# Light-Guided Surface Plasmonic Bubble Movement via Contact Line De-Pinning by In-Situ Deposited Plasmonic Nanoparticle Heating

Qiushi Zhang[1], Yunsong Pang[1], Jarrod Schiffbauer[1,2], Aleksandar Jemcov[1], Hsueh-Chia Chang[1,3], Eungkyu Lee[1]*, and Tengfei Luo[1,3,4]*

[1]Department of Aerospace and Mechanical Engineering, University of Notre Dame, Notre Dame, IN, USA

[2]Department of Physics, Colorado Mesa University, Grand Junction, CO, USA

[3]Department of Chemical and Biomolecular Engineering, University of Notre Dame, Notre Dame, USA.

[4]Center for Sustainable Energy of Notre Dame (ND Energy), University of Notre Dame, Notre Dame, USA.

*Correspondence to: elee18@nd.edu, and tluo@nd.edu

**ABSTRACT**

Precise spatio-temporal control of surface bubble movement can benefit a wide range of applications like high-throughput drug screening, combinatorial material development, microfluidic logic, colloidal and molecular assembly, etc. In this work, we demonstrate that surface bubbles on a solid surface are directed by a laser to move at high speeds (> 1.8 mm/s), and we elucidate the mechanism to be the de-pinning of the three-phase contact line (TPCL) by rapid plasmonic heating of nanoparticles (NPs) deposited in-situ during bubble movement. Based on our observations, we deduce a stick-slip mechanism based on asymmetric fore-aft plasmonic heating: local evaporation at the front TPCL due to plasmonic heating de-pins and extends the front TPCL, followed by the advancement of the trailing TPCL to resume a spherical bubble shape to minimize surface energy. The continuous TPCL drying during bubble movement also enables well-defined contact line deposition of NP clusters along the moving path. Our finding is beneficial to various microfluidics and pattern writing applications.

**INTRODUCTION**

The ability to manipulate bubbles in liquids promises to greatly advance robotic handling of liquid, which has a wide range of applications such as high throughput genomics screening, combinatorial material development, healthcare diagnosis, microstructure assembly, microfluidic logic, and vapor generation[1-8]. In a related field of liquid droplets manipulation, extensive research has been performed and different mechanisms have been thoroughly studied, such as electrostatic interaction, optical tweezers, and Marangoni effect[9-11]. In contrast, the fundamentals of controlled bubble movement on a solid surface have been under-explored. The Marangoni effect has been commonly cited as the key mechanism driving gas bubble movement in liquids. In 1888, Quincke[12] moved a surface air bubble in water using the solutal Marangoni effect enabled by an alcohol-induced surface tension gradient, which caused the bubble to migrate toward the alcohol-rich region (lower surface tension). Since surface tension is temperature-dependent, the Marangoni effect can also be realized by imposing a temperature gradient across the bubble, also called thermo-capillary convection. In 1959, Young et al.[13] leveraged such an effect to prevent the ascension of a bubble in a liquid column as the thermal Marangoni effect drives the bubble away from the colder end with higher surface tension.

To introduce a temperature gradient, the photothermal conversion process has been leveraged because a tightly focused light can precisely heat a specific location around a surface bubble and this allows more accurate manipulation of surface bubbles. There have been several demonstrations of using focused light to direct bubble movements on solid surfaces coated with optically resistive thin-films (e.g., metals, semiconductors, or metal oxides) [14-19]. In these studies, light is focused on the optically resistive thin-films to generate heat which creates a surface bubble, and various mechanisms have been proposed to manipulate the location of surface bubble. For example, Hu et al. [14,15] have shown that a 200-μm air bubble can be actuated by pre-defined light patterns focused on an amorphous silicon substrate that absorbs the light and generates heat. When the laser dislocates away from the surface bubble, it has been found that the surface bubble moves toward the new location of the laser spot. They attributed the bubble movement to the thermo-capillary flow generated around the bubble. In another two experiments, Fujii et al.[16] and Lin et al.[17] have used pre-deposited Au film and indium tin oxide film, respectively, and proposed that, as the laser spot moves, a new vapor bubble was generated at the current laser spot and the old bubble collapsed, and such high rate phenomena exhibited as if a bubble was moving continuously. On the other hand, another strategy has been proposed by Zheng et al.[18], Zhao et al.[19], and Zou et al.[20], where the de-pinning of the three-phase contact line (TPCL) of the surface bubble can trigger the bubble to move. In their experiments, the liquid at the front TPCL of a bubble can be rapidly evaporated by laser heating of the Au thin-film, leading to the de-pinning of the trailing TPCL and then the migration of the surface bubble toward the laser spot. While most of the above-discussed works use an absorbing layer on the surface to transduce light into heat, Armon et al.[21] have shown that surface bubbles generated in a metallic nanoparticle (NP) suspension can be directed by a laser spot without any optically resistive thin-films on the surface, which can usually degrade light transmission efficiency or require multiple fabricating processes in vacuum. In their discussion, they attributed the interesting phenomenon of laser-guided surface bubble movement to the thermo-capillary convective flow. However, the underlying mechanism has not been clearly explained or indisputably confirmed, and a detailed study is needed.

In this work, we present evidence showing that the thermal evaporation-induced de-pinning of the front TPCL triggers the surface bubble movement in a plasmonic NP suspension. In the NP-water suspension, thermo-capillary convection due to volumetric heating brings NPs to the TPCL, which then work as an intense heat source by plasmonic resonance to induce local evaporation to de-pin the front TPCL and extend it forward. This is followed by the advancement of the trailing TPCL in a sequential stick-slip mechanism involving the fore and aft positions of the bubble. During bubble translation, surface NP clusters are left behind by the de-pinned TPCL through contact line deposition. With pre-deposited NPs, bubble movement can reach a high speed of at least 1.8 mm/s. By comparing the stick-slip motion with the spatial distribution of the deposited Au NPs, we find that the bubble lags more on the Au NP-deficient region while it translates faster in the region with abundant Au NPs. Using high-speed videography with interferometry, we indeed observe that the front TPCL is pushed forward when the laser spot overlaps with the front contact line, which sequentially leads to the de-pinning of the trailing TPCL and eventually leads the bubble to slip forward within ~ 1 ms. Based on the interferometry, we find that the driving force to de-pin the trailing contact line is two order-of-magnitude larger than the force induced by the thermo-capillary convective flow surrounding the surface bubble. This confirms that the TPCL de-pinning due to the plasmonic NPs heating is the main reason for the laser-directed surface bubble movement. The results of this work hence elucidate the fundamental mechanism of laser-directed surface bubble movement in plasmonic NP suspensions. The possibility of high-precision bubble manipulation has useful practical implications for a wide range of microfluidic applications.

**RESULTS AND DISCUSSION**

The experimental setup to generate, move, and monitor the surface bubble is schematically illustrated in Fig. 1a. We disperse Au NPs (Nanospectra Bioscience, Inc) consisting of a silica core (~ 100 nm of diameter) and an Au shell (~ 10 nm of thickness) in deionized (DI) water and contain it in a quartz cuvette. A femtosecond pulsed laser (repetition rate of 80.7 MHz and pulse duration of 200 fs) with a Gaussian intensity profile with a $1/e^2$ radius of 20 μm and a center wavelength of 800 nm is directed to the cuvette and tightly focused on the interface between the suspension and the cuvette wall. The wavelength of the laser coincides with the surface plasmonic resonance (SPR) peak of the used Au NPs. The laser heats up the Au NPs and a surface bubble can be generated at the laser spot. The bubble is allowed to grow to a certain size (radius, 20 μm < $R$ < 50 μm), after which the laser spot starts to translate along the surface (y-direction, Fig. 1a) with a certain velocity ($v_{laser}$). We note that gravity is in the negative y-direction. A high-speed camera (NAC image technology, HX-7) is used to record the bubble generation and movement.

In experiments, it was observed that the generated surface bubble can follow the movement of the laser spot instantaneously and intimately (Movie S1), and Fig. 1b shows representative optical images of a moving bubble from the top view at an interval of 200 ms. For more detailed analyses, the bubble movement is also recorded from the side view at a fine time resolution of 0.2 ms (Fig. 2a and Movie S2). In Fig. 2a, it is clear that the bubble is attached to the quartz surface, where a reflection image of the bubble is seen. The laser beam passes through the surface bubble from the bottom in the z-direction. It is observed that the laser beam coming out of the top of the bubble is skewed towards the laser moving direction. Such a distorted beam shape resulted by the light refraction at the top surface of the bubble suggests that the laser beam slightly precedes the center of the bubble during laser and bubble movement. After careful observation of the refracted laser beam shape, we see a gradual spreading of the beam leading edge towards the laser moving direction before it abruptly retracts (see Figs. 2b and 2c and Movie S3). This implies that the laser beam moves away from the bubble center gradually (it is referred to as the "lag" motion in Fig. 2b) and then the bubble suddenly displaces to center at the new laser location (it is referred to as the "advance" motion in Fig. 2c), which suggests that the bubble moves in a lag-and-advance stick-slip manner.

In the NP-water suspension, the laser thermally excites the suspended NPs at the SPR, which leads to volumetric heating of the volume irradiated by the laser beam [22-26]. The volumetric heating induces a thermo-capillary convective flow as schematically shown in Fig. 3a. The flow can bring NPs in the suspension towards the TPCL of the surface bubble [27,28]. This is evident by tracking the movement of the glowing dots, where the glowing dots correspond to the scattered light from the plasmonic Au NPs. In Fig. 2a and Movie S2, we can clearly see that the glowing dots move towards the surface bubble (e.g., one dot indicated by the red arrows in Fig. 2a). By tracking the NP motion, we estimate an average flow speed of ~ 30 mm/s in the laser irradiated region above the surface bubble. We reproduced thermal convective flow using a finite element method (FEM) simulation by assuming volumetric heating (see Fig. 3b and Supporting Information, SI1). Figure 3b clearly shows that the induced flow direction is towards the surface bubble, which agrees with the migration direction of the NP in the experiments (Fig. 2a). In addition, the calculated flow velocity is on the same order of magnitude of the observed result (see the scale bar in Fig. 3b). This flow eventually brings the suspended NPs to the TPCL of the surface bubble [27,28]. As the liquid at the TPCL dries out by the laser heating, clusters of NPs are left on the surface as stains, which can be seen from Fig. 1b (black arrow). These immobilized NP clusters can serve as a heat source when subject to laser irradiation. However, the volumetric heating is found to be key to reproduce the experimentally observed thermo-capillary convective flow direction. If we assume surface heating to be dominated from the deposited NPs at the surface, the thermo-capillary convective flow would be in the opposite direction (see Fig. 3c) as predicted and observed by a number of previous studies [16-19], which is apparently not consistent with our experimental observation.

Although laser heating of the deposited NP clusters is not the main cause of the thermo-capillary convective flow, it is critical to bubble movement. Using the NP-water suspension, we vary the laser moving speed and find that the surface bubble can follow the laser instantaneously until the laser speed reaches 560 μm/s (Fig. 4a and see Movie S5). Given that the thermo-capillary convective flow has a much higher speed (30 mm/s) than the laser moving speed (< 1 mm/s), this indicates that the thermo-capillary convective flow is not the likely culprit for surface bubble motion. As shown later, the viscous stress and pressure acting on the surface bubble solely induced by the thermo-capillary convective flow is much smaller than the driving force needed to move the bubble. Instead, we find that the density of the NP clusters stain left on the solid surface due to contact line deposition steadily decreases as the bubble moving speed increases, as seen in the dark-field optical microscope (Fig. 4b) and SEM (Fig. 4c) images. Therefore, we propose that it is the result of a lack of Au NPs delivered to the TPCL that leads to too small a heating intensity to de-pin the TPCL, which makes the surface bubble to fail to follow the laser spot at high velocities. If so, this can be potentially overcome when the surface is pre-deposited with Au NP clusters. To confirm this, we create a path of the Au NP stain by generating and moving a surface bubble slowly ($v_{laser}$ = 100 μm/s). We then generate a new surface bubble and move the laser along the pre-deposited Au NP path with a constant acceleration of 3 mm/s$^2$ for a total travel distance of 1 mm. The bubble is able to follow the laser spot instantaneously for the whole process (see Fig. 4d and Movie S6) with speed up to 1800 μm/s. This confirms that the deposited Au NPs is responsible for the surface bubble movement. We should note that 1800 μm/s is the largest speed our translation stage can reach, and thus it should represent the lower limit of the achievable speed of the laser-directed bubble movement. We also note that in the Au NP stain, there are aggregated NPs like dimers or trimers. While the scattering peak of these aggregated NPs is likely to be red-shifted[29], we find that their optical absorption efficiencies are similar to (or even higher than) that of the single NPs (see Supporting Information, SI5). Thus, the heating and the resultant depinning effect by these aggregated NPs should also be similar to that by single NPs.

We also microscopically resolve a moving surface bubble from the side view with a time interval of 50 μs, which displays a very interesting lag-and-advance bubble motion. From the video, we track and analyze the travel distances of the surface bubble along the y-direction as a function of time. As seen in Fig. 5a, the distance traveled by the bubble in any instance is shorter than that of the laser spot. In addition, the movements of the surface bubbles are not continuous (see Movie S7), but are a series of lag-and-advance motions (see Fig. 5a). We also find that the lag-and-advance motion is in general correlated with the density of the deposited NP along the moving path (e.g., Fig. 5b). The lag state is prolonged when there are less NPs on the surface. When the NP density is low at the TPCL, the laser needs to move further so that the higher intensity portion of the Gaussian intensity profile overlaps with the lower NP density to generate sufficient heat to evaporate the fluid at the TPCL and de-pin it. We also note that the NP deposition during the bubble movement is stochastic (see Fig. 5c) and it is possible that when the deposited NP density is too low, especially when the laser moves too fast, the de-pinning cannot happen. This is why the bubble can fail to follow the laser spot as previously shown (Fig. 4a).

To further investigate the Au NP stain effect decoupled from the surrounding thermo-capillary convection in the suspension, we purged the NP-water suspension and filled the cuvette with DI water. After generating a surface bubble on the pre-deposited NP path, the laser is moved again with a constant acceleration of 3 mm/s$^2$ for a total travel distance of 1 mm, and it is seen that the bubble can follow the laser movement in the whole process (see Supporting Information, SI2). This additionally verifies that the bubble movement is driven by the deposited Au NPs stain since there should be very weak thermo-capillary convection on the surface bubble movement in the DI water. In addition, it is worth mentioning that the size of the bubble in DI water keeps decreasing during the movement since the bubble is being cooled by quartz substrate when moving to a new location, while that in NP suspension shows increasing radius. This should be related to the volumetric heating in NP-water suspension which helps the dissolved gas in water expel into surface bubble during the moving process[30]. After degassing the NP-suspension using a mechanical pump, the growth rate of the moving surface bubble in the suspension is significantly reduced in comparison to that of the pristine suspension (see Supporting Information, SI2).

From the above results, we have found that the laser heating of the deposited Au NP clusters is the key to the moving bubble and its lag-and-advance stick-slip motion. The mechanism of this stick-slip motion is illustrated in Fig. 6a. Before the laser beam moves, the TPCL of a surface bubble is pinned with an equilibrium contact angle ($\theta_e$). When the laser beam moves forward slightly, the laser overlaps with the front contact line and heat up the deposited NP clusters there. The heating locally evaporates the liquid microlayer at the TPCL, pushing the contact line outward. This is also described as the "recoil force" [31-33] from the rapid evaporation of the liquid at the TPCL. As the front TPCL is pushed outward, the effect of the vapor/water surface tension will result in a contact angle larger than the equilibrium one. The trailing TPCL will then also possess the similar contact angle as the bubble minimizes the vapor/water surface energy. In the meantime, the trailing contact line is still pinned, and the bubble movement lags behind that of the laser. As the front TPCL is further extended following the laser movement, the contact angles further increase until a critical angle ($\theta_c$) is reached. Beyond this point, the pinning effect can no longer hold the trailing TPCL[34-36], it then retracts, and the whole bubble advances forward to follow the laser beam. Due to the pinning effect, the laser beam center will precede the center of the lagged bubble, leading to the asymmetric refraction of the beam coming out of the top of the bubble as previously discussed in Fig. 2a and 2b.

To obtain more insights into the stick-slip motion and visualize the propagation of the TPCL during surface bubble movement, we employ a laser interferometry setup similar to Ref. [20] to quantify the relative motion of the laser and the TPCL (see Supporting Information, SI3 for detail). The constructive and destructive patterns of a coherent light source (i.e., interference fringe patterns) in the microlayer under the surface bubble allows the identification of the TPCL. Figure 6b shows the laser interferometry images corresponding to each stage described in Fig. 6a. The full laser interferometry video recording the stick-slip motion of a surface bubble is provided in Movie S8. Figure 6c illustrates the distance between the laser spot center and the bubble center as a function of time with a time resolution of 0.1 ms. At first, in stages (i) and (ii), the bubble lags behind the moving laser spot, and the distance between two centers increases gradually. Then, in stage (iii) the laser spot overlaps with the front TPCL and push it forward because of heating up of the deposited NP clusters at the contact line. The laser spot keeps drying the contact line and pushing it to result in a contact angle larger than $\theta_c$. Finally, in stage (iv), after the pinning force can no longer hold the surface bubble, the bubble slips forward to "catch up" the laser spot. One point to mention here is that this "catch-up" motion of the surface bubble is extremely fast, which is finished within ~ 1 ms (Fig. 6c).

The interference fringe patterns also allow us to estimate the contact angle on the trailing TPCL. In the interferometry images, the distance between two neighboring constructive rings (dashed lines in Fig. 6d) in the radial direction ($\Delta d$) can determine the contact angle ($\theta$) via the following relation[37]:

$$\theta = \operatorname{atan}\left(\frac{\lambda}{2n\Delta d}\right) \qquad (1)$$

where $\lambda$ is the vacuum wavelength of the coherent light, and $n$ is the refractive index of water ($n = 1.33$). Using this relation, we can estimate the contact angles at the trailing TPCL to be $\theta_e \sim 11°$ at stage (i), and $\theta \sim 24°$ at stage (iii). Two representative interferometry images in Fig. 6d clearly shows the changes of fringe patterns from the equilibrium (i.e., stage (i)) to stage (iii), where it is evident that the shape of the TPCL is changed to an oval shape from a circle (solid white lines in Fig. 6d) as the laser spot overlaps with the front end of the TPCL. In addition, our calculated contact angles match well with the reported values due to the TPCL de-pinning process of surface bubble on a hydrophilic $SiO_2$ surface[20], which uses optically resistive thin-films buried under the $SiO_2$ surface to induce the TPCL de-pinning.

The increased trailing contact angle at stage (iii) means that Young's equation will yield a non-zero net force, as the projected liquid-vapor surface tension at the trailing TPCL is weakened due to the increased contact angle. Here, it is reasonable to assume that the surface bubble can only move when this non-zero net force is larger than the pinning force that holds the surface bubble. Using Young's equation, the net force ($F_{net}$) at the trailing TPCL can be expressed as (see Supporting Information, SI4 for detail):

$$F_{net} = r_{TPCL}\gamma_{LG}\left(2\cos\theta_e + \int_{\pi}^{2\pi}\cos\theta(\phi)\sin\phi\,d\phi\right), \qquad (2)$$

where $r_{TPCL}$ is the radius of the TPCL ($r_{TPCL} = 33$ μm) of the surface bubble (white solid line in Fig. 6d), $\gamma_{LG}$ is the water-air surface tension ($\gamma_{LG} = 72$ mN m$^{-1}$), $\phi$ is the azimuthal angle on the surface plane, $\theta(\phi)$ is the contact angle depending on $\phi$ at stage (iii). We assume that $\theta(\phi)$ is the equilibrium angle at $\phi = \pi$, and it linearly increases to 24° at $\phi = 2\pi/3$, and then linearly decreases to the equilibrium angle at $\phi = 2\pi$. We expect the assumption of the linear relation between the contact angle and the azimuthal angle to give the correct order of magnitude in force estimation. This leads to $F_{net} \sim 1.8 \times 10^{-7}$ N according to Eq. (2), which is the minimum force needed to de-pin the TPCL and allow the surface bubble to displace. We further compare this pinning force with force from the viscous stress and pressure induced by the thermo-capillary convective flow from the volumetric heating of the NP suspension. According to our calculation (see Supporting Information, SI1), it is found that the force on the surface bubble by the thermo-capillary convective flow is $\sim 5 \times 10^{-9}$ N when the laser spot overlaps with the front contact line. This is almost two orders of magnitude lower than the estimated pinning force. This reasonably leaves the front TPCL de-pinning due to plasmonic heating as the main reason for the laser directed surface bubble movement. We note that if the surface is super-hydrophilic, the pinning force will be smaller and thus the surface bubble may move faster as directed by the laser.

Finally, as a potential application, we leverage our finding to merge two surface bubbles. The surface bubble merging process is particularly important to chemical reactants or catalyst delivery on surface[38,39]. Using a laser, we first generate a surface bubble with a radius of ~ 120 µm (referred to as the "target bubble") on the quartz surface in the suspension. We then create another surface bubble with a radius of ~ 30 µm (referred to as the "carrier bubble") at a remote location away from the target bubble and move the laser spot to deliver the carrier bubble toward the target bubble to let them merge together. In Movie S4, it is clearly seen that the larger target bubble absorbs the smaller carrier bubble instantaneously as their vapor/liquid interfaces contact each other. The merging process occurs at a time scale of less than 200 µs, which is due to the Ostwald ripening effect[40,41]. We believe that the laser-guided merging process of two surface bubbles can enable potential applications beyond the demonstrated ones such as micro-pattern writting[16-18] and micro-particle assembly[14,19].

**CONCLUSIONS**

In conclusion, we present evidence showing that the surface bubble movement in an Au NP-water suspension is triggered by the thermal evaporation-induced de-pinning of the front TPCL, followed by the advancement of the trailing TPCL. The thermo-capillary convection brings NPs to the TPCL, which then works as a heat source to induce local evaporation to de-pin the TPCL and thus move the bubble. Meanwhile, NP clusters are deposited on the surface due to TPCL drying. Along the line-written path of pre-deposited NPs, bubble movement can reach high speeds of at least 1.8 mm/s. High-speed videography and the analysis of the diffracted laser light of the microlayer near TPCL both show that the bubble moves in a stick-slip manner while the laser translates continuously. The interferometry confirms the front contact line extension by the laser-NP heating, the de-pinning process of trailing TPCL followed by the slip of the surface bubble. Evaluating the driving force at the trailing TPCL due to the increased contact angle confirm that the thermal Marangoni effect has an insignificant role in the laser-directed surface bubble movement. Not only do the results of this work help elucidate the fundamental physics of laser-directed surface bubble movement in a NP suspension, but also, they demonstrate the capability for controlled contact line deposition and precise control of bubble movement without pre-deposited optically resistive thin-films. There can be useful implications for a wide range of microfluidics and directed-assembly applications[42,43].

**METHODS**

**Sample Preparation:** A quartz cuvette (Hellma, Sigma-Aldrich, 10 mm x 10 mm) with 4 windows was sequentially cleaned with DI water, acetone, and isopropyl alcohol in an ultrasonic bath, and then baked at 150 °C for 10 mins to evaporate organic solvents before use. The cleaned quartz cuvette was filled with Au NPs suspension, where the Au NPs (AuroShell, Nanospectra Biosciences, Inc.) have a spherical shape with a radius of ~ 60 nm and consists of a silica core (radius of ~ 50 nm) and an Au shell (thickness of ~ 10 nm), dispersed in DI water at a concentration of 2 x $10^{15}$ particles/m$^3$. The gas-poor NP-water suspension was obtained by ~ 4 h degassing. The concentration of air in each case was quantized by measuring the concentration of oxygen in the liquid. The concentrations of oxygens are ~ 8.4 mg/L for the pristine NPs-water suspension and ~ 4.3 mg/L for the gas-poor suspension.

**Characterization of Surface Bubble Dynamics:** In the experimental setup schematically shown in Fig. 1a, a Ti:sapphire crystal in an optical cavity (Spectra Physics, Tsunami) emitted the mode-locked monochromatic pulsed laser. The laser has a center wavelength of 800 nm, full-width-half-maximum length of 10 nm, power between 485 ~ 705 mW, the pulse duration of ~ 200 fs and the repetition rate of 80.7 MHz. The laser beam was focused in the suspension in the cuvette through an objective lens (10x for Fig.1 to Fig. 5 and 20x for Fig. 6, Edmund Optics). A white LED with 300 lm was used for the illumination source. The illumination source passes through the sample to enters another objective lens (20x, Edmund Optics) and is then focused on a digital camera (HX-7, NAC), which is appropriately positioned to record the top view or the side view of the surface bubble. The cuvette was mounted on a motorized translational stage (MT1-Z8, Thorlabs). An optical shutter was positioned in front of the pulsed laser. The motorized stage, the optical shutter, and the digital camera were electrically connected to a digital controller (KDC101, Thorlabs). To record the formation and motion of the surface bubble, the controller was controlled by pre-defined sequential parameters through a LabVIEW interface. The recorded movies in the digital camera were analyzed using a customized image processing software in MATLAB. In the interferometry experiment, we use a coherent laser source with a wavelength of 632.8 nm (HeNe, 2 mW, Thorlabs). The detail experimental setup is shown in the Supporting Information, SI2.

**ACKNOWLEDGEMENTS**

This work is supported by National Science Foundation (1706039) and the Center for the Advancement of Science in Space (GA-2018-268). T.L. would also like to thank the support from the Dorini Family endowed professorship in energy studies.

## AUTHOR CONTRIBUTIONS

E.L. Q.Z. and T.L. designed the experiments, and E.L. and Q. Z. set up the experiments. E.L. and Q.Z. performed the experiment. E.L., J.S., Y. P. and A.J. designed the simulations and E.L. performed the simulations. E.L., Q.Z., J.S., A. J., H.C.C., and T.L. discussed the results and the mechanism of bubble movements. E.L., Q.Z., and T.L. wrote the manuscript, J.S., Y. P., A. J., and H.C.C. revised it.

## COMPETING INTERESTS

The authors declare no conflict of interest.

## REFERENCES

1. Whiteside, G. M. The Origin and the Future of Microfluidics. *Nature* **2006**, *442*, 368-373.
2. DeMello, A. J. Control and Detection of Chemical Reaction in Microfluidic Systems. *Nature* **2006**, *442*, 394-402.
3. Song, H.; Chen, D. L.; Ismagilov, R. F. Reactions in Droplets in Microfluidic Channels. *Angew. Chem. Int. Ed.* **2006**, *45*, 7336-7356.
4. Prakash, M.; Gershenfeld, N. Microfluidic Bubble Logic. *Science* **2007**, *315*, 832-835.
5. Chung, S. K.; Cho, S. K. On-Chip Manipulation of Objects Using Mobile Oscillating Bubbles. *J. Micromech. Microeng.* **2008**, *18*, 125024.

6. Neumann, O.; Urban, A. S.; Day, J.; Lal, S.; Nordlander, P.; Halas, N. J. Solar Vapor Generation Enabled by Nanoparticles. *ACS Nano* **2013**, *7*(1), 42-49.

7. Fang, Z.; Zhen, Y.; Neumann, O.; Polman, A.; de Abajo, F. J. G.; Nordlander, P.; Halas, N. J. Evolution of Light-Induced Vapor Generation at a Liquid-Immersed Metallic Nanoparticle. *Nano Lett.* **2013**, *13*(4), 1736-1742.

8. Dongare, P. D.; Alabastri, A.; Neumann, O.; Nordlander, P.; Halas, N. J. Solar Thermal Desalination as a Nonlinear Optical Process. *Proc. Natl. Acad. Sci. U. S. A.* **2019**, *116* (27), 13182-13187.

9. Squires, T. M.; Quake, S. R. Microfluidics: Fluid Physics at the Nanoliter Scale, *Rev. Mod. Phys.* **2005**, *77*, 977.

10. Baigl, D. Photo-Actuation of Liquids for Light-Driven Microfluidics: State of the Art and Perspectives. *Lab Chip* **2012**, *12*, 3637-3653.

11. Nelson, W. C.; Kim, C. J. Droplet Actuation by Electrowetting-on-Dielectric (EWOD): A Review. *J. Adhesion Sci. Technol.* **2012**, *26*, 1747-1771.

12. Quincke, G. Ueber Periodische Ausbreitung an Flussigkeitsoberflachen und Dadurch Hervorgerufene Bewegungserscheinungen. *Ann. Phys. (Berl.)* **1888**, *271*, 580-642.

13. Young, N. O.; Goldstein, J. S.; Block, M. J. The Motion of Bubbles in a Vertical Temperature Gradient. *J. Fluid Mech.* **1959**, *6*, 350-356.

14. Hu, W.; Ishii, K. S.; Ohta, A. T. Micro-Assembly Using Optically Controlled Bubble Microbots. *Appl. Phys. Lett.* **2011**, *99*, 094103.

15. Hu, W.; Ishii, K. S.; Fan, Q.; Ohta, A. T. Hydrogel Microbots Actuated by Optically Generated Vapour Bubbles. *Lab Chip* **2012**, *12*, 3821-3826.

16. Fujii, S.; Fukano, R.; Hayami, Y.; Ozawa, H.; Muneyuki, E.; Kitamura, N.; Haga, M. Simultaneous Formation and Spatial Patterning of ZnO on ITO Surfaces by Local Laser-Induced Generation of Microbubbles in Aqueous Solution of [Zn(NH3)4]2+. *ACS Appl. Mater. Interfaces* **2017**, *9*, 8413-8419.

17. Lin, L.; Peng, X.; Mao, Z.; Li, W.; Yogeesh, M. N.; Rajeeva, B. B.; Perillo, E. P.; Dunn, A. K.; Akinwande, D.; Zheng, Y. Bubble-Pen Lithography. *Nano Lett.* **2016**, *16*, 701-708.

18. Zheng, Y.; Liu, H.; Wang, Y.; Zhu, C.; Wang, S.; Cao, J.; Zhu, S. Accumulating Microparticles and Direct-Writing Micropatterns Using a Continuous-Wave Laser-Induced Vapor Bubble. *Lab Chip* **2011**, *11*, 3816-3820.

19. Zhao, C.; Xie, Y.; Mao, Z.; Zhao, Y.; Rufo, J.; Yang, S.; Guo, F.; Mai, J. D.; Huang, T. J. Theory and Experiment on Particle Trapping and Manipulation via Optothermally Generated Bubbles. *Lab Chip* **2014**, *14*, 384-391.

20. Zou, A.; Gupta, M.; Maroo, S. C. Origin, Evolution, and Movement of Microlayer in Pool Boiling. *J. Phys. Chem. Lett.* **2018**, *9*, 3863-3869.

21. Armon, N.; Greenberg, E.; Layani, M.; Rosen, Y. S.; Magdassi, S.; Shpaisman, H. Continuous Nanoparticle Assembly by a Modulated Photo-Induced Microbubble for Fabrication of Micrometric Conductive Patterns. *ACS Appl. Mater. Interfaces* **2017**, *9*, 44214-44221.

22. Lukianova-Hleb, E.; Hu, Y.; Latterini, L.; Tarpani, L.; Lee, S.; Drezek, R. A.; Hafner, J. H.; Lapotko, D. O. Plasmonic Nanobubbles as Transient Vapor Nanobubbles Generated around Plasmonic Nanoparticles. *ACS Nano* **2010**, *4*, 2109-2123.

23. Boulais, E.; Lachaine, R.; Meunier, M. Plasma Mediated Off-Resonance Plasmonic Enhanced Ultrafast Laser-Induced Nanocavitation. *Nano Lett.* **2012**, *12*, 4763-4769.

24. Lukianova-Hleb, E.; Volkov, A. N.; Lapotko, D. O. Laser Pulse Duration is Critical for Generation of Plasmonic Nanobubbles. *Langmuir* **2014**, *30*, 7425-7434.

25. Metwally, K.; Mensah, S.; Baffou, G. Fluence Threshold for Photothermal Bubble Generation Using Plasmonic Nanoparticles. *J. Phys. Chem. C* **2015**, *119*, 28586-28596.

26. Lachaine, R.; Boutopoulos, C.; Lajoie, P.-Y.; Boulais, E.; Meunier, M. Rational Design of Plasmonic Nanoparticles for Enhanced Cavitation and Cell Perforation. *Nano Lett.* **2016**, *16*, 3187-3194.

27. Fujii, S.; Kanaizuka, K.; Toyabe, S.; Kobayashi, K.; Muneyuki, E.; Haga, M. Fabrication and Placement of a Ring Structure of Nanoparticles by a Laser-Induced Micronanobubble on a Gold Surface. *Langmuir* **2011**, *27*, 8605-8610.

28. Uwada, T.; Fujii, S.; Sugiyama, T.; Usman, A.; Miura, A.; Masuhara, H.; Kanaizuka, K.; Haga, M. Glycine Crystallization in Solution by cw Laser-Induced Microbubble on Gold Thin Film Surface. *ACS Appl. Mater. Interfaces* **2012**, *4*, 1158-1163.

29. Tira, C.; Tira, D.; Simon, T.; Astilean, S. Finite-Difference Time-Domain (FDTD) Design of Gold Nanoparticle Chains with Specific Surface Plasmon Resonance. *J. Mol. Struct.* **2014**, *1072*, 137-143.

30. Wang, Y.; Zaytsev, M. E.; The, H. L.; Eijkel, J. C. T.; Zandvliet, H. J. W.; Zhang, X.; Lohse, D. Vapor and Gas-Bubble Growth Dynamics around Laser-Irradiated, Water-Immersed Plasmonic Nanoparticles. *ACS Nano* **2017**, *11*, 2045-2051.

31. Nikolayev, V. S.; Beysens, D. A. Boiling Crisis and Non-Equilibrium Drying Transition. *Europhys. Lett.* **1999**, *47* (3), 345-351.

32. Lagubeau, G.; Merrer, M. L.; Clanet, C.; Quere, D. Leidenfrost on a Ratchet. *Nat. Phys.* **2011**, *7*, 395-398.

33. Nikolayev, V. S.; Chatin, D.; Garrabos, Y.; Beysens, D. A. Experimental Evidence of the Vapor Recoil Mechanism in the Boiling Crisis. *Phys. Rev. Lett.* **2006**, *97*, 184503.

34. Karapetas, G.; Sahu, C. K.; Matar, O. K. Effect of Contact Line Dynamics on Thermocapillary Motion of a Droplet on Inclined Plate. *Langmuir* **2013**, *29*, 8892-8906.

35. Li, Q.; Zhou, P.; Yan, H. J. Pinning-Depinning Mechanism of Contact Line During Evaporation on Chemically Patterned Surfaces: a Lattice Boltzmann Study. *Langmuir* **2016**, *32*, 9389-9396.

36. Mohammadi, M.; Sharp, K. V. The Role of Contact Line (Pinning) Force on Bubble Blockage in Microchannels. *J. Fluid. Eng.* **2015**, *137*, 0312081-0312087.

37. Voutsinos, C. M.; Judd, R. L. Laser Interferometric Investigation of the Microlayer Evaporation Phenomenon. *J. Heat Transfer* **1975**, *97* (1), 88-92.

38. Adleman, J. R.; Boyd, D. A.; Goodwin, D. G.; Psaltis, D. Heterogenous Catalysis Mediated by Plasmon Heating. *Nano Lett.* **2009**, *9*, 4417-4423.

39. Sirsi, S.; Borden, M. Microbubble Compositions, Properties and Biomedical Applications. *Bubble Sci Eng Technol.* **2009**, *1*, 3-17.

40. Binder, K. Theory for the Dynamics of "Clusters." II. Critical Diffusion in Binary Systems and the Kinetics of Phase Separation. *Phys. Rev. B* **1977**, *15*, 4425.

41. Watanabe, H.; Suzuki, M.; Inaoka, H.; Ito, N. Ostwald Ripening in Multiple-Bubble Nuclei. *J. Chem. Phys.* **2014**, *141*, 234703.

42. Darhuber, A. A.; Troian, S. M. Principle of Microfluidic Actuation by Modulation of Surface Stresses. *Annu. Rev. Fluid. Mech.* **2005**, *37*, 425-455.

43. Duan, C.; Wang, W.; Xie, Q. Review Article: Fabrication of Nanofluidic Devices. *Biomicrofluidics* **2013**, *7*, 026501.

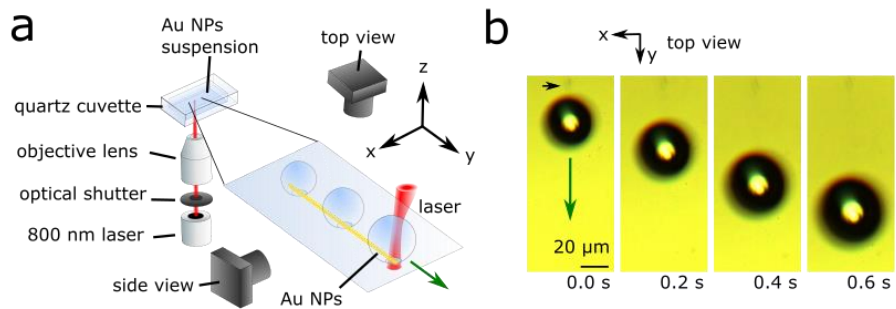

**Figure 1.** Characterization of moving surface bubbles. (a) Schematic experimental setup to characterize the motion of the surface bubble. (b) Optical images from the top view of the moving bubble on the quartz substrate in the NP suspension guided by the laser with a velocity ($v_\text{laser}$) of 100 μm/s and a power ($P_{laser}$) of 550 mW. The green arrows depict the direction of the laser translation. In (b), the black arrow depicts the formed Au NPs stain on the path of the moving bubble.

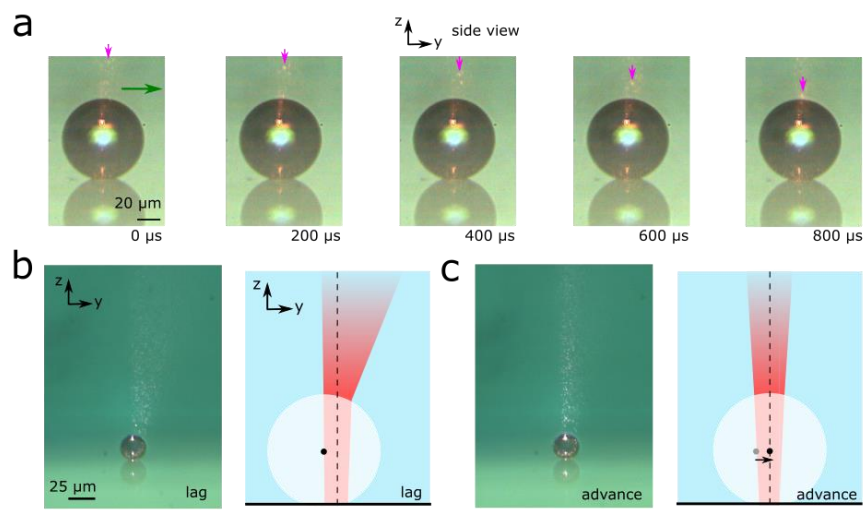

**Figure 2.** (a) Optical images from the side view of the moving bubble on the quartz substrate in the NP suspension guided by the laser with a velocity ($v_{laser}$) of 100 μm/s and a power ($P_{laser}$) of 550 mW. In (a), the magenta arrow indicates the scattered laser light from nano-bubbles with Au NPs in the suspension, which propagates towards the top of the surface bubble. (b and c) Refracted laser beam passing out of the top surface of the bubble by optical imaging (left) and schematic illustration (right) in (b) the 'lag' state and (c) the 'advance' state of the bubble movement.

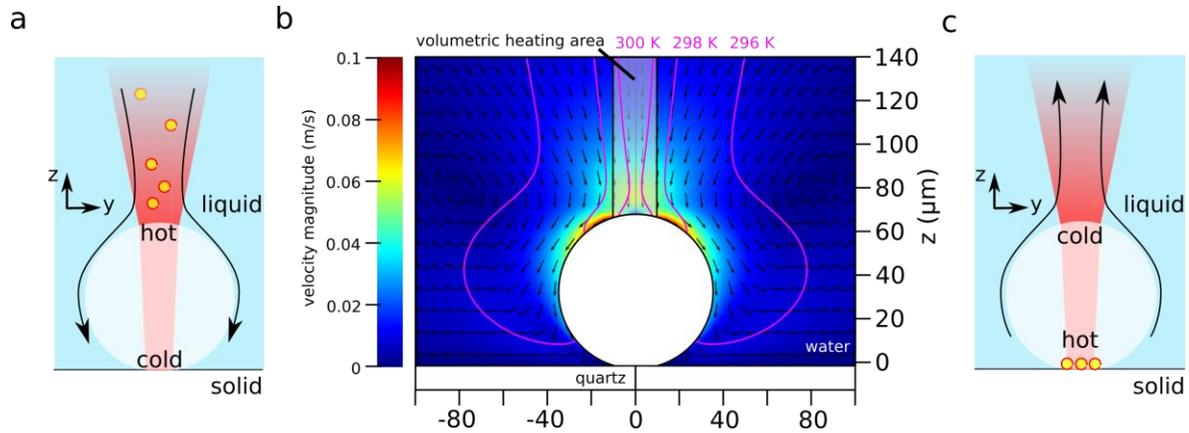

**Figure 3.** Thermo-capillary convective flow surrounding the surface bubble. (a) Schematic of the vertical thermo-capillary convective flow direction when the suspension is subject to volumetric heating in the laser beam covered region. (b) Calculated thermo-capillary convective flow around the surface bubble when the laser induces volumetric heating in the suspension. The shade area in the suspension depicts the volumetric heating region covered by the laser irradiation. The black arrows show the direction of the convective flow. The magenta solid lines are the isothermal contours of temperature. (c) Schematic of the thermo-capillary flow when the heat source is located at the surface of the substrate.

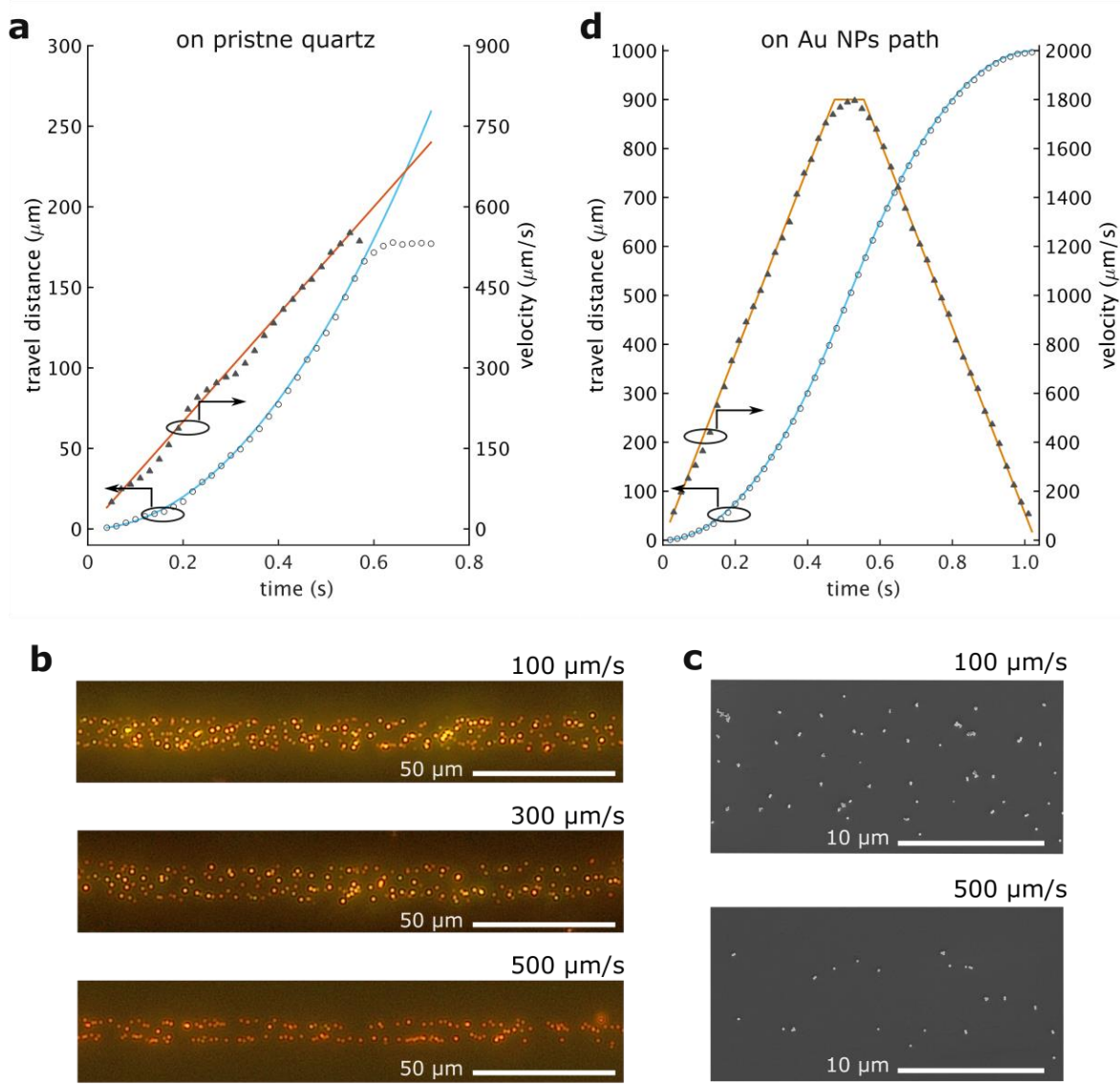

**Figure 4.** Displacement, velocity, and the deposited Au NPs of moving surface bubbles on surfaces and the effect of the pre-deposited Au NPs. (a) The travel distance and velocity of the laser (solid lines) and the surface bubble (symbols) as a function of time in the NP suspension on pristine quartz surface. The laser moves with a constant acceleration (or de-acceleration) of $a_{laser.} = 1$ mm/s$^2$ (b) Dark-field optical microscope image of the Au NPs stain deposited along the path of the moving surface bubble with $v_{laser} = 100$ μm/s (top), 300 μm/s (middle), and 500 μm/s (bottom). (c) Scanning electron microscope images of the deposited Au NPs stains from $v_{laser} = 100$ μm/s (top)

and 500 μm/s (bottom). (d) The travel distance and velocity of the laser (solid lines) and the surface bubble (symbols) as a function of time in the NP suspension with pre-deposited NP stain on the quartz surface; here $a_{laser} = 3$ mm/s$^2$.

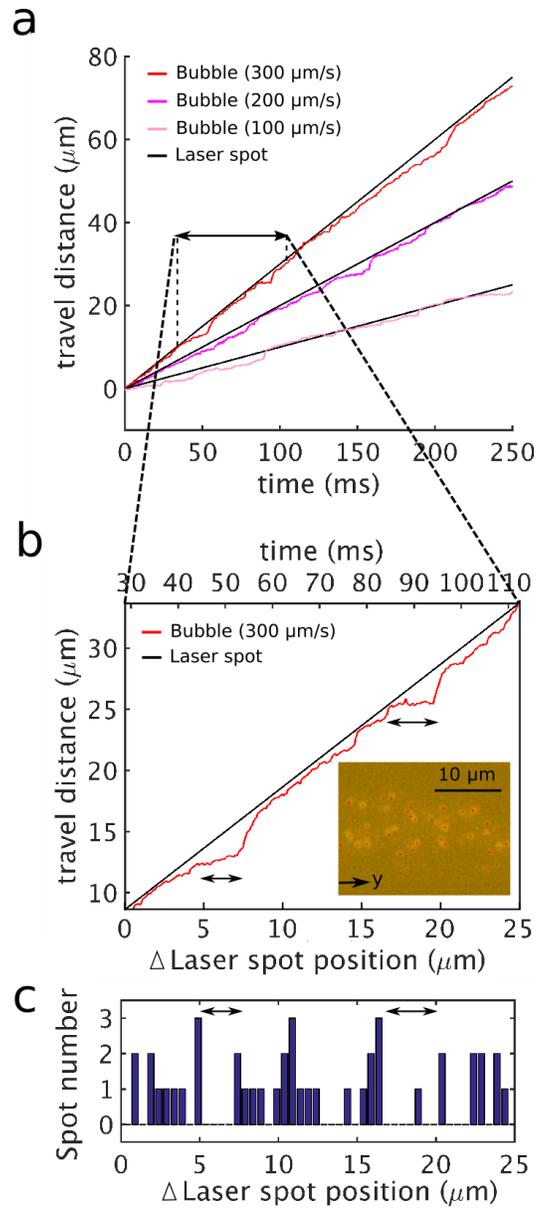

**Figure 5.** Stick-slip motion of the surface bubble. (a) Travel distance of the surface bubble (color lines) and the laser (black lines) as a function of time when the laser moves with $v_{laser}$ = 100, 200, and 300 μm/s. (b) Travel distance of the surface bubble as a function of time corresponding to the time period indicated by the arrow in (a). The inset shows the optical image of deposited Au NPs on the corresponding travel path. The bottom axis shows the relative laser spot position in the y-

direction. (c) Deposited Au NPs spot density along the laser moving path, which corresponds to the inset optical image in (b).

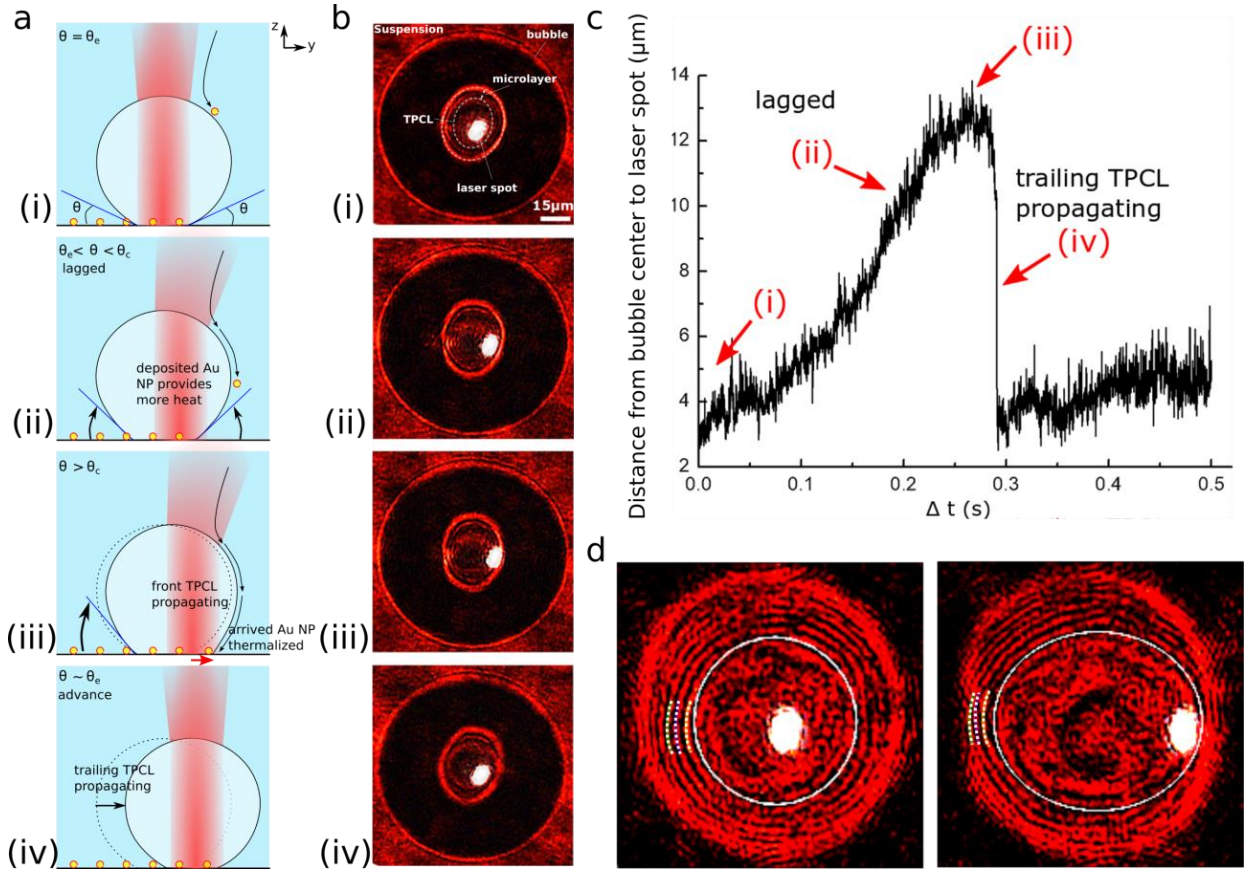

**Figure 6.** De-pinning of contact line in Lag-and-advance motion of the surface bubble. (a) Schematic illustration of the lag-and-advance of the surface bubble. (i) The contact lines are pinned with the equilibrated contact angle ($\theta = \theta_e$) before the laser beam moves. (ii) When the laser moves slightly forward and the beam starts to overlap with the deposited Au NPs around the front TPCL, the NPs provide more heat for water evaporation at the contact line. This pushes the TPCL outward and leads to an increase of the front contact angle. To minimize the vapor/liquid surface tension, the trailing contact angle increases accordingly. The increased trailing contact angle is still smaller than a critical angle ($\theta_e < \theta < \theta_c$), and the trailing TPCL is still pinned. In this phase, bubble lags behind the translating laser spot. (iii) The laser continues moving forward, and the front contact line is further pushed outward (red arrow). This process eventually increases the trailing contact angle to reach critical angle, and (iv) finally, the trailing contact line overcomes the pinning effect and moves forward, which enables the surface bubble to advance forward. (b) Optical

interferometry images of a surface bubble in lag-and-advance motion. Each stage from (i) to (iv) in (b) corresponds to that in (a). The brighter white dots show the locations of laser spot. Here, $P_{laser}$ is 500 mW and 20x Objective lens is used to focus the laser. The light source for the interferometry has the wavelength of 630 nm and power of 2 mW. (c) The distance between the center of bubble and that of the laser spot as a function of time. The red arrows indicate the time corresponding to each stage in (b). (d) Optical interferometry images at stage (i) (left) and stage (iii) (right). The white solid lines indicate the TPCL and the area inside the white solid lines is the dry-out region. The periodic red and black rings outside the white solid lines correspond to the fringe patterns of coherent light source in microlayers, respectively. The white dotted lines indicate the first-three constructive interference rings on the side of the trailing contact line.

# Supporting Information for

# Light-Guided Surface Plasmonic Bubble Movement via Contact Line De-Pinning by In-Situ Deposited Plasmonic Nanoparticle Heating

Qiushi Zhang[1], Yunsong Pang[1], Jarrod Schiffbauer[1,2], Aleksandar Jemcov[1], Hsueh-Chia Chang[1,3], Eungkyu Lee[1]*, and Tengfei Luo[1,3,4]*

[1]Department of Aerospace and Mechanical Engineering, University of Notre Dame, Notre Dame, IN, USA

[2]Department of Physics, Colorado Mesa University, Grand Junction, CO, USA

[3]Department of Chemical and Biomolecular Engineering, University of Notre Dame, Notre Dame, USA.

[4]Center for Sustainable Energy of Notre Dame (ND Energy), University of Notre Dame, Notre Dame, USA.

*Correspondence to: elee18@nd.edu, and tluo@nd.edu

## SI1. Calculating thermo-capillary convective flow surrounding the surface bubble

To calculate the thermo-capillary convective flow around the surface bubble by the volumetric heating of the plasmonic Au NP suspension, we assume the following six conditions. (1) The liquid flow and heat transfer are at steady state. (2) In the liquid water, the flow is laminar and incompressible without body forces (note that gravity is perpendicular to the observed flow and thus is not considered to be important here), which satisfies the following momentum equation:

$$\rho(\vec{u} \cdot \nabla)\vec{u} - \nabla \cdot (\mu(\overrightarrow{\nabla u} + \overrightarrow{\nabla u}^{\mathrm{T}}) - p\overleftrightarrow{I}) = 0 \quad \text{(s1)}$$

, and continuity equation:

$$\rho(\nabla \cdot \vec{u}) = 0 \quad \text{(s2)}$$

where $\rho$ is the density of water, $\mu$ is the dynamic viscosity of water, $\vec{u}$ is the velocity vector, $p$ is the pressure, and $\overleftrightarrow{I}$ is a 3x3 identity tensor. (3) The gas medium inside the surface bubble and quartz considered as non-fluidic rigid materials, i.e., we are effectively modeling a single-phase liquid flow around a rigid bubble. (4) The volumetric heating of the Au NP suspension is the only heat source, which supplies the heat to the liquid water with the following heat transfer equations: In water,

$$\rho C_p \vec{u} \cdot \nabla T - k_w \nabla^2 T = Q_v \quad \text{(s3)}$$

where $C_p$ is the heat capacity of water at constant pressure, $T$ is the temperature, $k_w$ is the thermal conductivity of water, and $Q_v$ is the heat generation density by the volumetric heating, and in the medium of gas and quartz,

$$-k_s \nabla T = q \tag{s4}$$

where $k_s$ is the thermal conductivity of gas (or quartz), and $q$ is the heat flux coming through the gas/water boundary (or gas/quartz boundary). (5) The plasmonic heating of Au NPs in water is related to the optical attenuation factor ($\alpha = 262$ m$^{-1}$) as follows:

$$Q_v = \eta_{abs} \alpha \frac{P_{laser}}{2\pi\sigma^2} \exp\left[-\left(\frac{(y-d)^2}{2\sigma^2} + \alpha(z-h)\right)\right] \tag{s5}$$

where the optical attenuation factor of the suspension is extracted from the absorbance spectrum to represent the volumetric-averaged light extinction of the Au NPs in water, $P_{laser}$ is the power of laser, $\sigma = 11$ μm for the 10x objective lens (or 3 μm for the 20x objective lens) is the width of the Gaussian laser spot determined from a beam profiler and the recorded optical image of the surface bubble, $d$ is the distance between the center of the bubble and the center of the laser spot, $h$ is the height of the bubble surface from the quartz surface at the distance $d$ from the center, and $\eta_{abs}$ is the optical absorption efficiency of NPs, which is 0.2 determined by the ratio of the absorption quality factor and the extinction quality factor of the Au NP in water. (6) Finally, the surface of the bubble (gas/water boundary) has a slip boundary condition with the Marangoni effect as:

$$\left[\mu(\overrightarrow{\nabla u} + \overrightarrow{\nabla u}^T) - \left(p + \frac{2}{3}\mu(\nabla \cdot \vec{u})\right)\overleftrightarrow{I}\right]\hat{n} = \gamma \nabla_t T \tag{s6}$$

where $\hat{n}$ is the normal outward vector to the surface of the bubble, $\gamma$ is the temperature derivative of the water/gas surface tension, and $\nabla_t$ is the gradient of the tangent vector to the surface of the bubble.

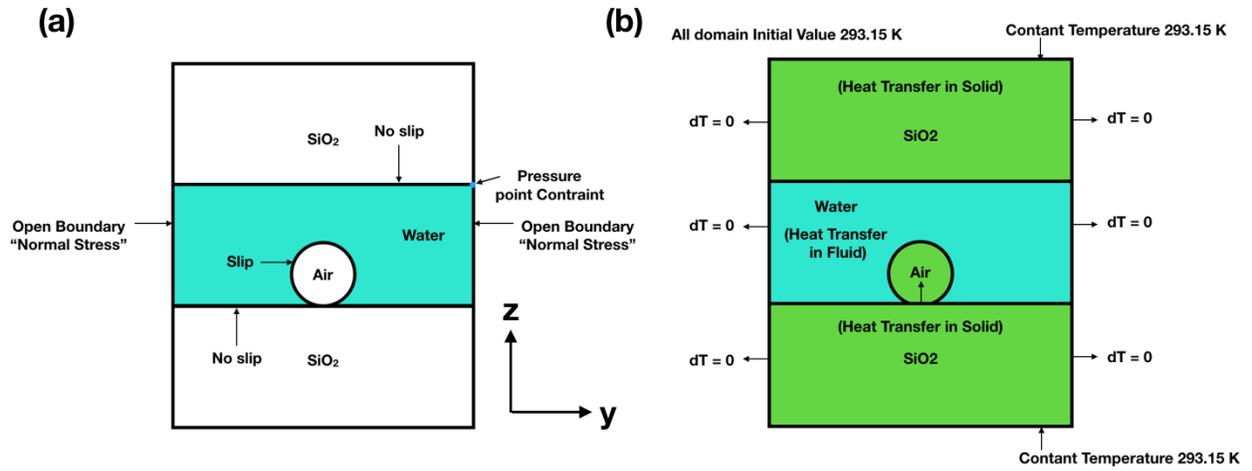

**Figure S1.** Schematic structure and boundary conditions for the calculation of thermo-capillary convective flow surrounding the surface bubble on quartz. (a) Boundary conditions for the laminar flow calculation in water. (b) Boundary conditions for the heat transfer calculation in the system.

Figure S1 shows the schematic structure and physical boundary conditions of the simulation system, where a surface bubble in NP-water suspension is sandwiched by two quartz plates. This mimics the experimental structure. We use a finite element method (FEM, COMSOL Multiphysics) to solve $\vec{u}$, $p$, and $T$ profiles using equations (s1) to (s6). The domain of the system is divided into $2 \times 10^6$ finite elements. For results shown in Fig. 2b of the main text, the radius of a bubble is 35 μm, the objective lens is 10x, $d$ is 0 μm, and $P_{\text{laser}}$ is 550 mW. For the discussion with the interferometry images in the main text, the radius of a bubble is 120 µm, the objective lens is 20x, $d$ is from 0 µm to 33 µm, and $P_{laser}$ is 500 mW. After the FEM simulation, we calculate the y-component of the force ($F_{th,y}$) acting on the surface bubble by the thermo-capillary convective flow using the following formula:

$$F_{th,y} = |\int^{A}\left[\mu\left(\left(\frac{\partial v}{\partial x}+\frac{\partial u}{\partial y}\right)\hat{x} + \left(2\frac{\partial v}{\partial y}\right)\hat{y} + \left(\frac{\partial v}{\partial z}+\frac{\partial w}{\partial y}\right)\hat{z}\right) - p\hat{y}\right]\cdot\hat{n}\mathrm{d}a\,| \quad (s7)$$

where $\hat{x}$, $\hat{y}$, and $\hat{z}$ are the unit vectors in the Cartesian coordinate and $A$ is the interfacial area of the surface bubble for the water/gas boundary.

## SI2. Surface bubble growth during movement

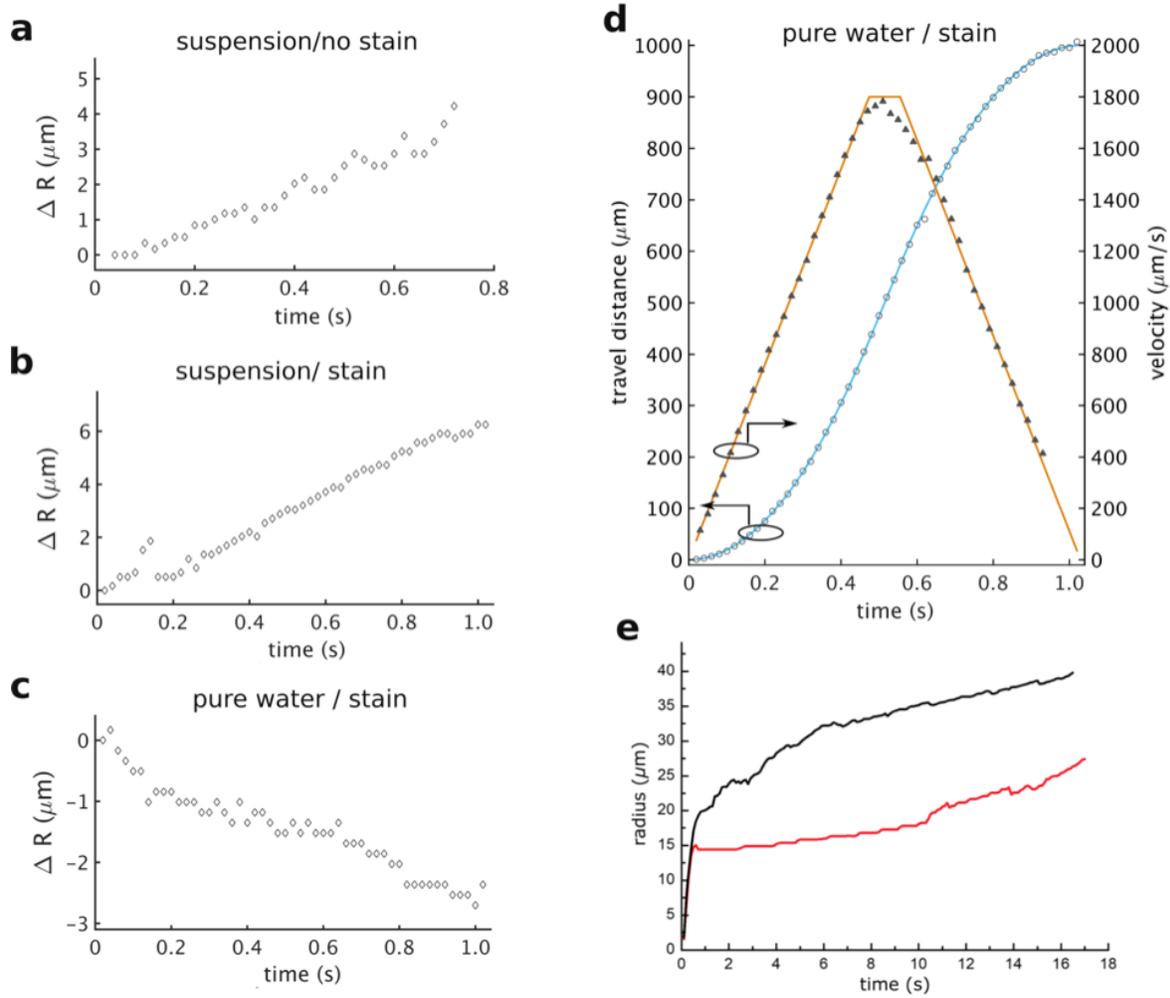

**Figure S2.** (a to c) Radius change ($\Delta R$) of the moving surface bubbles as a function of time in three cases: (a) in the NP suspension without pre-deposited NP stain on the quartz surface; (b) in the NP suspension with pre-deposited NP stain on the surface; and (c) in the deionized water with pre-deposited NP stain. (d) The travel distance and velocity of the laser (solid lines) and the surface bubble (symbols) as a function of time in case (c), laser spot acceleration $a_{laser} = 3$ mm/s$^2$. (e) Radius change ($\Delta R$) of the surface bubbles as a function of time with (red) or without (black) degassing process.

**SI3. Experimental setup of high-speed videography with laser interferometry**

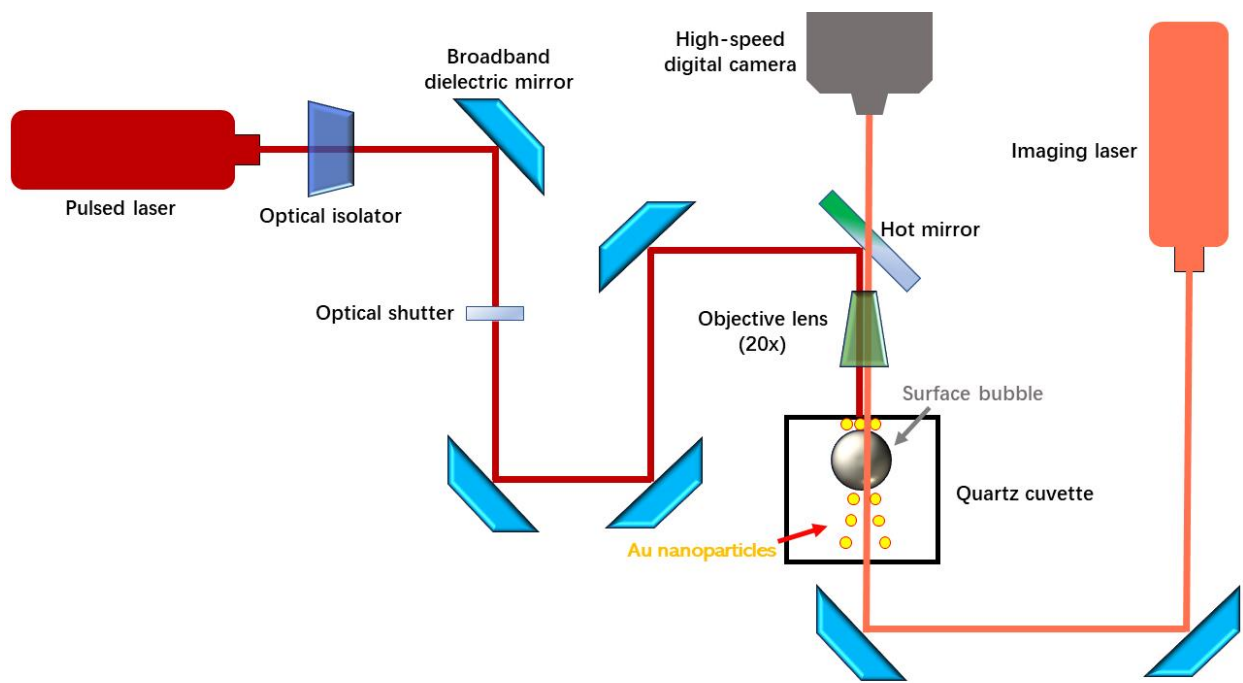

**Figure S3.** Schematic of the experimental setup of high-speed videography with laser interferometry.

## SI4. Calculating the net force at the trailing TPCL

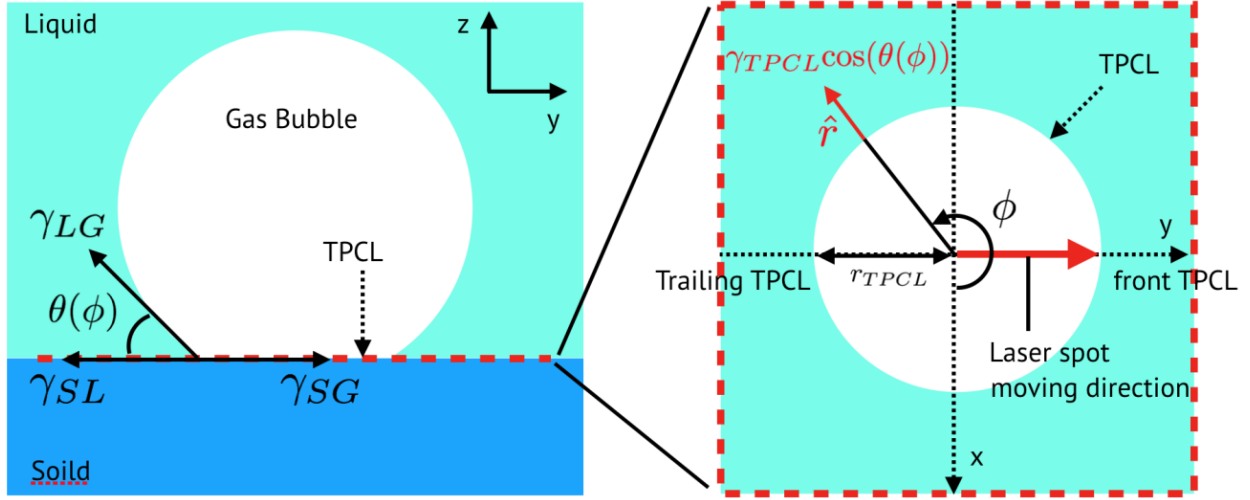

**Figure S4.** Schematic of geometrical configuration for calculating the net force at the trailing TPCL.

In the equilibrium system of the surface bubble in liquid, Young's equation at the TPCL at the azimuthal angle $\phi$ can be given by (see Figure S4 for the geometrical configuration):

$$\gamma_{SL}\hat{r} + \gamma_{LG}\cos\theta(\phi)\hat{r} + \gamma_{SG}(-\hat{r}) = 0 \qquad (s8)$$

where the equilibrium system depicts the center of the laser spot is at the center of the bubble (corresponding to the stage (i) in the main text), $\gamma_{SL}$ is the surface tension at the solid-liquid interfaces, $\gamma_{SG}$ is the surface tension at the solid-gas interfaces, $\gamma_{LG}$ is the surface tension at the liquid-gas interfaces, $\hat{r}$ is the unit radial vector on the x-y plane where the TPCL is on, and $\theta(\phi)$ is the contact angle at the azimuthal angle of $\phi$, which is defined as Figure S4. Here, $\theta(\phi)$ is $\theta_e \sim 11°$ for all $\phi$ in the equilibrium system, which gives:

$$\gamma_{SL}\hat{r} + \gamma_{SG}(-\hat{r}) = -\gamma_{LG}\cos\theta_e\,\hat{r}. \tag{s9}$$

As the laser spot moves along the y-direction and is overlapped with the front TPCL (corresponding to the stage (iii) in the main text), $\theta(\phi)$ is increased around the trailing TPCL as shown in Figure 5d in the main text. In this case, Young's equation can yield a non-zero net surface tension force ($f_{net}(\phi)$) at a certain $\phi$ and we can re-write equation (s8) using (s9) as:

$$f_{net}(\phi) = \gamma_{LG}\cos\theta(\phi)\,\hat{r} - \gamma_{LG}\cos\theta_e\,\hat{r} \neq 0 \tag{s10}$$

where the direction of $f_{net}(\phi)$ is the negative $\hat{r}$ since $\theta_e < \theta(\phi)$ for at the stage (iii). By integrating $f_{net}$ along the trailing TPCL ($\pi < \phi < 2\pi$), we can evaluate the net force ($F_{net}$) at the trailing TPCL as:

$$F_{net} = \int_{\pi}^{2\pi} f_{net}(\phi) r_{TPCL} \,d\phi \tag{s9}$$

$$= r_{TPCL}\int_{\pi}^{2\pi}(\gamma_{LG}\cos\theta(\phi)\sin\phi\,\hat{y} - \gamma_{LG}\cos\theta_e\sin\phi\,\hat{y})d\phi$$

$$= r_{TPCL}\gamma_{LG}\left(\int_{\pi}^{2\pi}\cos\theta(\phi)\sin\phi\,d\phi + 2\cos\theta_e\right)\hat{y}.$$

## SI5. Calculating the absorption cross section of single Au NP, Au NPs dimer and Au NPs trimer

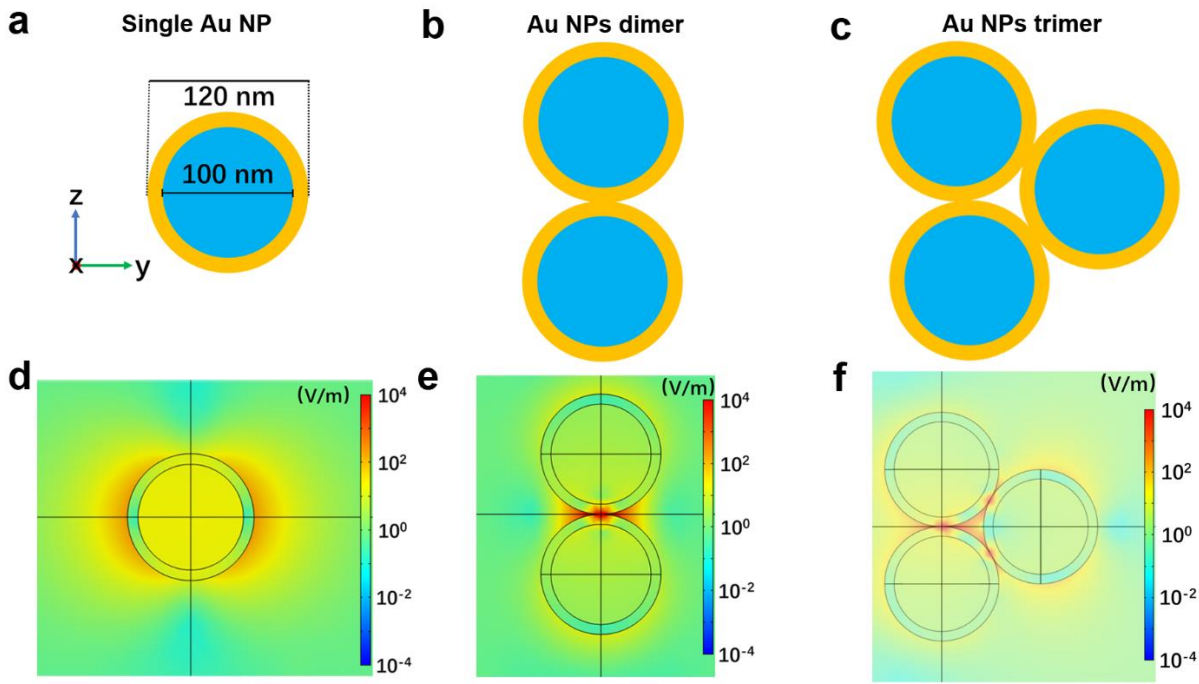

**Figure S5.** Schematics of the optical configurations of single Au NP (a), Au NPs dimer (b) and Au NPs trimer (c), when a linear z-polarized plane wave is incident into the NPs. Here, the incident light has a wavelength of 800 nm and propagates along the x-axis. The size of each Au NP (120 nm diameter) in the three cases is same. The absolute amplitude of the electric field profiles are shown for single Au NP (d), Au NPs dimer (e) and Au NPs trimer (f).

During the movement of a surface bubble, Au NPs are deposited in-situ along the path (as shown in Figs. 4b and 4c). Some of these deposited Au NPs are found to aggregate together into structures, like dimer or trimer. To compare the photothermal conversion efficiencies of single Au NP, Au NPs dimer and Au NPs trimer, we performed full-wave electromagnetic calculations with finite element method to estimate the light absorption cross section in each case. In the results seen in Table S1, we can find that dimer or timer can have the absorption cross-section that is similar to (or even higher than) that of the single Au NP.

|  | Single NP | NPs Dimer | NPs Trimer |
|---|---|---|---|
| **Polarized in y-axis** | $2.3 \times 10^{-14}$ | $1.8 \times 10^{-14}$ | $3.7 \times 10^{-14}$ |
| **Polarized in z-axis** | $2.3 \times 10^{-14}$ | $2.9 \times 10^{-14}$ | $3.8 \times 10^{-14}$ |

**Table S1.** The calculated absorption cross section ($m^2$) in the cases of single Au NP, Au NPs dimer and Au NPs trimer.